\documentclass[12pt]{iopart}
\usepackage{iopams,graphicx}
\usepackage{amstext,wasysym}
\usepackage[square,numbers,sort&compress]{natbib}
\usepackage{longtable}

\makeatletter
\newcommand{\nuc}[2]{\ensuremath{^{\text{#1}}\text{#2}}}
\makeatother

\begin{document}

\review[Current Status and Future Potential of Nuclide Discoveries]{Current Status and Future Potential of Nuclide Discoveries}

\author{M. Thoennessen}

\address{National Superconducting Cyclotron Laboratory and \\ Department of Physics \& Astronomy, \\ Michigan State University, East Lansing, MI 48824, USA}
\ead{thoennessen@nscl.msu.edu}
\begin{abstract}
Presently about 3000 different nuclei are known with about another 3000-4000 predicted to exist. A review of the discovery of  the nuclei, the present status and the possibilities for future discoveries are presented.
\end{abstract}

\maketitle

\section{Introduction}
\label{sec:intro}

The strong force, responsible for the binding of nucleons is one of the fundamental forces. In order to understand this force it is critical to know which combination of neutrons and protons can form a bound nuclear system. Even now, after more than 100 years of nuclear physics research this information is only known for the lightest elements. Thus the search for new nuclides with more and more extreme neutron to proton ratios continues to be important. The discovery of new nuclides also is the first step in exploring and measuring any properties of these nuclides.

Over the years more and more sophisticated detectors and powerful accelerators were developed to push the limit of nuclear knowledge further and further. At the present time about 3000 nuclides are known. Recently it was calculated that about 7000 nuclides are bound with respect to neutron or proton emission \cite{2012Erl01}. In addition, there are neutron and proton unbound nuclides which can have significantly shorter lifetimes or appear only for a very short time as a resonance. The properties of these nuclides beyond the ``driplines'' can also be studied with special techniques \cite{2012Bau01,2012Pfu01} and they are especially interesting because they represent the extreme limits for each element.

The present review gives a brief historical overview followed by a summary of the present status and a discussion of future perspectives for the discovery of new nuclides. Throughout the article the word nuclide is used rather than the widely used but technically incorrect term isotope. The term isotope is only appropriate when referring to a nuclide of a specific element.

\section{Historical Overview}
\label{sec:history}

It can be argued that the field of nuclear physics began with the discovery of radioactivity by Becquerel in 1896 \cite{1896Bec01} who observed the radioactive decay of what was later determined to be $^{238}$U \cite{1923Sod01,1931Ast02}. Subsequently, polonium ($^{210}$Po \cite{1898Cur03}), and radium ($^{226}$Ra \cite{1898Cur02}) were observed as emitting radioactivity, before Rutherford discovered the radioactive decay law and determined the half-life of radon ($^{220}$Rn \cite{1900Rut01}). He was also the first to propose the radioactive decay chains and the connections between the different active substances \cite{1905Rut01} as well as the identification of the $\alpha$-particle: ``...we may conclude that an $\alpha$-particle is a helium atom, or, to be more precise, the $\alpha$-particle, after it has lost its positive charge, is a helium atom'' \cite{1908Rut01}.

The distinction of different isotopes for a given element was discovered only in 1913 independently by Fajans \cite{1913Faj03} and Soddy \cite{1913Sod01} explaining the relationship of the radioactive chains. Soddy coined the name ``isotope'' from the Greek words ``isos'' (same) and ``topos'' (place) meaning that two different ``isotopes'' occupy the same position in the periodic table \cite{1913Sod02}.

The first clear identification of two isotopes of an element other than in the radioactive decay chains was reported by Thomson in 1913 using the positive-ray method: ``There can, therefore, I think, be little doubt that what has been called neon is not a simple gas but a mixture of two gases, one of which has an atomic weight about 20 and the other about 22'' \cite{1913Tho01}.

\begin{figure}
	\centering
	\includegraphics[scale=.8]{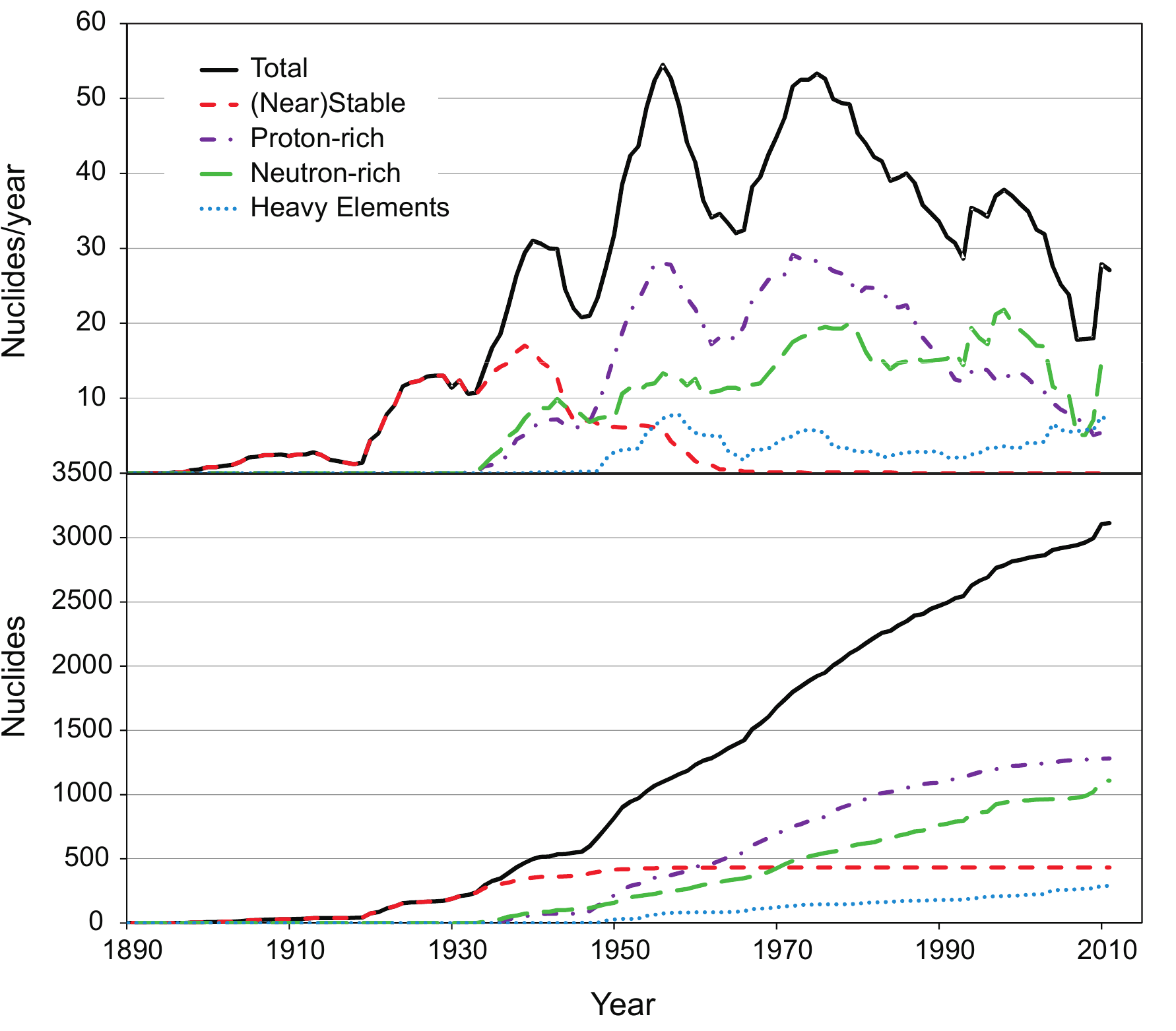}
	\caption{Discovery of nuclides as a function of year. The top figure shows the 10-year running average of the number of nuclides discovered per year while the bottom figure shows the cumulative number.  The total number of nuclides shown by the black, solid lines are plotted separately for near-stable (red, short-dashed lines), neutron-deficient (purple, dot-dashed lines), neutron-rich (green, long-dashed lines) and transuranium (blue, dotted lines) nuclides (see text for explanation). }
\label{f:by-year}
\end{figure}

Since this first step, continuous innovations of new experimental techniques utilizing the new knowledge gained about nuclides led to the discovery of additional nuclides. This drive to discover more and more exotic nuclides has moved the field forward up to the present day. Figure \ref{f:by-year} demonstrates this development where the number of nuclides discovered per year (top) and the integral number of discovered nuclides (bottom) are shown. In addition to the total number of nuclides (black, solid lines) the figure also shows the number of near-stable (red, short-dashed lines), neutron-deficient (purple, dot-dashed lines), neutron-rich (green, long-dashed lines) and transuranium (blue, dotted lines) nuclides. Near-stable nuclides are all nuclides between the most neutron-deficient and neutron-rich stable isotopes of a given element. Lighter and heavier radioactive isotopes of the elements are then classified as neutron-deficient and neutron-rich, respectively.

The figure shows that the rate of discovery was not smooth and the peaks can be directly related to the development of new experimental techniques as explained in the next subsections.

\subsection{Mass spectroscopy of stable nuclides}

In 1908, Rutherford and Geiger had identified the $\alpha$-particle as helium \cite{1908Rut01} and in 1913 Thompson accepted in addition to neon with mass number 20 the presence of a separate neon substance with mass number 22 which represented the beginning of mass spectroscopic methods to identify isotopes as separate identities of the same element with different mass numbers \cite{1913Tho01}. The first ``mass spectra'' were measured by Aston when he added focussing elements to his first ``positive ray spectrograph'' in 1919 \cite{1919Ast02}. From 1919 to 1930 the number of known identified nuclides jumped from 40 to about 200 mostly due to Aston's work. The development of more sophisticated mass spectrographs by Aston \cite{1927Ast03,1930Ast01} and others \cite{1932Bai02,1935Dem01,1937Nie02} led to the discovery of essentially most of the stable nuclides \cite{1942Ast01}.

\subsection{Nuclear reactions and first accelerators}

In 1919 Rutherford discovered nuclear transmutation: ``From the results so far obtained it is difficult to avoid the conclusion that the long-range atoms arising from collision of $\alpha$ particles with nitrogen are not nitrogen atoms but probably atoms of hydrogen, or atoms of mass 2'' \cite{1919Rut01}. He apparently observed the reaction $^{14}$N($\alpha$,p), however, it took six years before Blackett identified the reaction residue as the new nuclide $^{17}$O \cite{1925Bla01}. It took another seven years before in 1932 the discovery of the neutron by Chatwick \cite{1932Cha01} and the first successful construction of a particle accelerator by Cockcroft and Walton \cite{1932Coc03} led to the production of many new nuclides by nuclear reactions.

Cockcroft and Walton were able to prove the production of $^8$Be using their accelerator: ``...the lithium isotope of mass 7 occasionally captures a proton and the resulting nucleus of mass 8 breaks into two $\alpha$-particles...'' \cite{1932Coc01}; Harkins, Gans and Newson produced the first new nuclide ($^{16}$N) induced by neutrons ($^{19}$F(n,$\alpha$)) \cite{1933Har01} and in 1934, I. Curie and F. Joliot observed artificially produced radioactivity ($^{13}$N and $^{30}$P)\footnote{They also reported another activity assigned to $^{27}$Si, however, most likely they observed $^{28}$Al} in ($\alpha$,n) reactions for the first time \cite{1934Cur01}.

Also in 1934, Fermi claimed the discovery of a transuranium element in the neutron bombardment of uranium \cite{1934Fer02}. Although the possibility of fission was immediately mentioned by Noddack: ``It is conceivable that [...] these nuclei decay into several larger pieces'' \cite{1934Nod01}, even with mounting evidence in further experiments, Meitner, Hahn, and Strassmann did not take this step: ``These results are hard to understand within the current understanding of nuclei.'' \cite{1937Mei01} and ``As chemists we should rename Ra, Ac, Th to Ba, La, Ce. As `nuclear chemists' close to physics, we cannot take this step, because it contradicts all present knowledge of nuclear physics.'' \cite{1939Hah02}. After Meitner and Frisch correctly interpreted the data as fission in 1939 \cite{1939Mei01}, Hahn and Strassmann identified $^{140}$Ba \cite{1939Hah01} in the neutron induced fission of uranium. The first transuranium nuclide ($^{239}$Np) was then discovered a year later by McMillan and Abelson in neutron capture reactions on $^{238}$U \cite{1940McM01}.

Light particle induced reactions using $\alpha$-sources, neutron irradiation, fission, and continuously improved particle accelerators expanded the chart of nuclei towards more neutron-deficient, neutron-rich, and further transuranium nuclides for the next two decades. The number of nuclides produced every year continued to increase only interrupted by World War II. By 1950 the existing methods had reached their limits and the number of new isotopes began to drop. New technical developments were necessary to reach isotopes further removed from stability.

\subsection{Heavy-ion fusion evaporation reactions}

Although Alvarez demonstrated already in 1940 that it was possible to accelerate ions heavier than helium in the Berkeley 37-inch cyclotron \cite{1940Alv02}, the next major breakthrough came in 1950 when Miller {\it et al.} successfully accelerated detectable intensities of completely stripped carbon nuclei in the Berkeley 60-inch cyclotron \cite{1950Mil01}. Less than two months later Ghiorso {\it et al.} reported the discovery of $^{246}$Cf in the heavy-ion fusion evaporation reaction $^{238}$U($^{12}$C,4n) \cite{1951Ghi02}. This represented the first correct identification of a californium nuclide because the discovery of the element californium claimed the observation of $^{244}$Cf \cite{1950Tho07} which was later reassigned to $^{245}$Cf \cite{1956Che02}.

With continuous increases of beam energies and intensities fusion-evaporation reactions became the dominant tool to populate and study neutron-deficient nuclei. The peak in the overall production rate of new nuclides around 1960 is predominantly due to the production of new neutron-deficient nuclides and new super-heavy elements. Fusion-evaporation reactions are presently still the only way to produce super-heavy elements. The discovery of new elements relies on even further improvements in beam intensities and innovations in detector technology.

\subsection{Target and projectile fragmentation}

The significant beam energy increases of light-ion as well as heavy-ion accelerators opened up new ways to expand the nuclear chart. In the spallation or fragmentation of a uranium target bombarded with 5.3-GeV protons, Poskanzer {\it et al.} were able to identify several new neutron-rich light isotopes for the first time ($^{11}$Li, $^{12}$Be, and $^{14,15}$B) in 1966 \cite{1966Pos01}. Target fragmentation reactions were effectively utilized to produce new neutron-rich nuclides (see for example Ref. \cite{1969Han01}) using the ISOL (Isotope Separation On-Line) method. This technique was developed already 15 years earlier for fission of uranium by Kofoed-Hansen and Nielsen who discovered $^{90}$Kr and $^{90,91}$Rb \cite{1951Kof01} .

The inverse reaction, the fragmentation of heavy projectiles on light-mass targets was successfully applied to produce new nuclides for the first time in 1979 by bombarding a beryllium target with 205~MeV/nucleon $^{40}$Ar ions \cite{1979Sym01}. Projectile fragmentation began to dominate the production of especially neutron-rich nuclei starting in the late 1980s when dedicated fragment separators came online. For an overview of the various facilities, for example the LISE3 spectrometer at GANIL \cite{1991Mue01}, the RIPS separator at RIKEN \cite{1992Kub01}, the A1200 and A1900 separators at NSCL \cite{1991She01,2003Mor03}, and the FRS device at GSI \cite{1992Gei01} see Ref. \cite{1998Mor01}. In addition to these separators a significant number of nuclides were discovered at storage rings, see for example Refs. \cite{2008Fra01,2010Che01}.

The most recent increase in the production rate of new nuclides is predominantly due to new technical advances at GSI \cite{2010Alv01,2010Che01,2011Jan01} and the new next generation radioactive beam facility RIBF \cite{2007Yan01} with the separator BigRIPS \cite{2008Ohn01} at RIKEN.

\subsection{Discoveries of isotopes, isotones, and isobars}

\begin{figure}
	\centering
	\includegraphics[scale=.9]{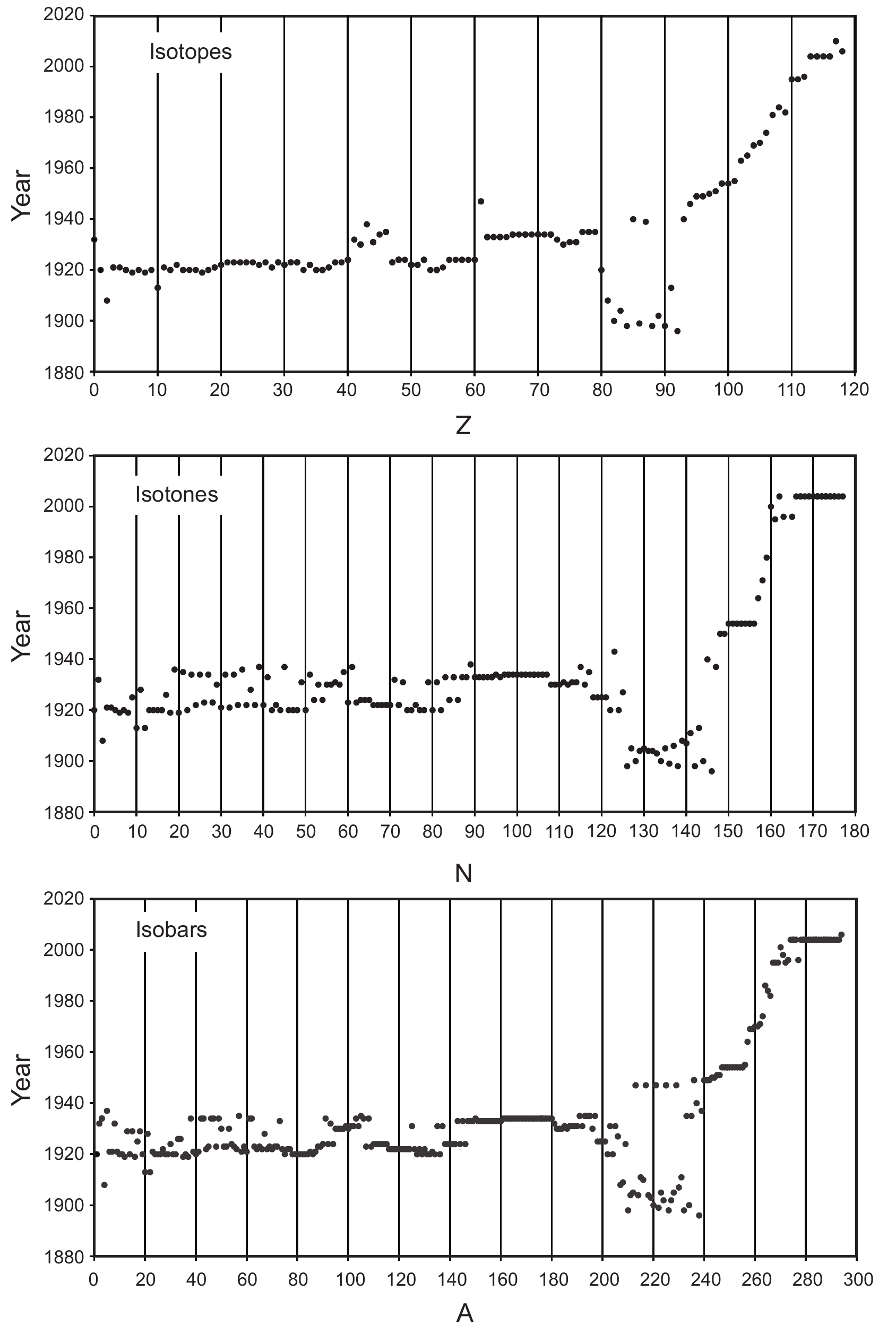}
	\caption{Discoveries of isotopes (top), isotones (middle), and isobars (bottom).}
\label{f:iii}
\end{figure}

It is interesting to follow the discovery of nuclides over the years as a function of isotopes (Z = constant), isotones (N = constant) and isobars (A = constant) as shown in the top, middle, and bottom panels of Figure \ref{f:iii}, respectively.

Unique characteristics of isotopes of elements from the radioactive decay chains were determined around 1900, and although the concept of isotopes was not established at that time these observations can be taken as the first identification of isotopes of these elements.
For most of the elements up to Z = 60 the first isotope was discovered in the early 1920s with exception of the transition metals of the 5$^{th}$ period between niobium and palladium which were identified for the first time in the 1930s. Also, as mentioned earlier, isotopes of helium ($^4$He or the $\alpha$-particle  \cite{1908Rut01}) and neon ($^{20,22}$Ne \cite{1913Tho01} were discovered earlier and the neutron was discovered in 1932 \cite{1932Cha01}.

Isotopes of the remaining stable elements were identified by the late 1930s. The last four missing elements below uranium were discovered by the identification of their specific isotopes. They were technetium (Z = 43) in 1938 \cite{1938Seg01}, francium (Z = 87) in 1939 \cite{1939Per01}, astatine (Z = 85) in 1940 \cite{1940Cor01}, and promethium (Z = 61) in 1947 \cite{1947Mar01}. Transuranium elements were then discovered starting in 1940 with the identification of neptunium ($^{239}$Np) \cite{1940McM01} at an approximately constant rate of about one element every three years (also see Figure \ref{f:seaborg-elements}).

Plotting the year of discovery as a function of isotones reveals another pattern. In the light mass region -- approximately between chlorine and zirconium (N $\sim$ 20 -- 50) -- the even-N isotones were discovered around 1920 while it took about another 15 years before the odd-N isotones were identified. This is due to the significantly smaller abundances of the even-Z/odd-N isotones in this mass region. In contrast, the abundances are more equally distributed in the lanthanide region (N $\sim$ 80 -- 110). While the advances in the discovery of new elements was fairly constant, the discovery of isotones displays a different pattern.

Although intense neutron irradiation of plutonium in the Idaho Materials Test Reactor did not discover any new elements, the successive neutron capture reactions produced many new isotones. In 1954 alone seven new isotones (N = 150 -- 156) were discovered. However, in the following 40 years only one additional isotone was added per decade.

At Dubna hot fusion reactions were used to populate new elements leading to the discovery of 15 new isotones within one year (2004) up to the heaviest currently known isotone of N = 177. The recent discovery of element 117 and 118 did not push the isotone limit any further. It should be mentioned that the isotone N = 164 has not yet been identified (see also Section \ref{subsec:heavy}).

The pattern of the discovery as a function of mass number up to A $\sim$ 200 shown in the bottom panel of Figure \ref{f:iii} mirrors approximately the pattern of the isotones.  Until 1937, when Meitner, Hahn, and Strassmann \cite{1937Mei01} discovered $^{239}$U, the discovery of radioactivity by Becquerel in 1896 \cite{1896Bec01} later attributed to $^{238}$U represented the heaviest nuclide. The missing (4n+3) radioactive decay chain observed in 1943 by Hagemann et al. \cite{1947Hag01} filled in the gaps at masses 213, 217, 221, 225, and 229. Currently the heaviest element (Z = 118) also represents the heaviest nuclide (A = 294).

\section{Current Status}
\label{sec:status}

Recently a comprehensive overview of the discovery of all nuclides was completed \cite{2012Tho01}. Details of the discovery of 3067 nuclides were described in a series of articles beginning in 2009 \cite{2009Gin01} with the latest ones being currently published. During this time another 38 nuclides were discovered for a total of 3105 nuclides observed by the end of 2011.  Table \ref{t:elements} lists the total number and the range of currently known isotopes for each element. It should be mentioned that for some elements not all isotopes between the most neutron-deficient and the most neutron-rich isotopes have been observed. In light neutron-rich nuclei these are $^{21}$C, $^{30}$F, $^{33}$Ne, $^{36}$Na, and $^{39}$Mg. The cases in the neutron-deficient medium-mass and the superheavy mass region are discussed in Sections \ref{subsec:proton} and \ref{subsec:heavy}, respectively. The table also lists the year of the first and most recent discovery as well as the reference for the detailed documentation of the discovery.

While the recognition for the discovery of a new element is well established with strict criteria set by the International Union of Pure and Applied Chemistry (IUPAC) and the International Union of Pure and Applied Physics (IUPAP) \cite{1976Har02,1991IUP01} the discovery of the different isotopes for a given element is not well defined \cite{2004Tho01}. The nuclides included in Table \ref{t:elements} had to be (1) clearly identified, either through decay-curves and relationships to other known nuclides, particle or $\gamma$-ray spectra, or unique mass and element identification, and (2) published in a refereed journal. In order to avoid setting an arbitrary lifetime limit for the definition of the existence of a nuclide, particle-unbound nuclides with only short-lived resonance states were included. Isomers were not considered separate nuclides.


{\setlength{\LTleft}{0pt}
\setlength{\LTright}{0pt}

\setlength{\tabcolsep}{0.5\tabcolsep}

\renewcommand{\arraystretch}{1.0}
\renewcommand{\baselinestretch}{1.0}

\footnotesize 
\begin{longtable}{@{\extracolsep\fill}lrrrcrrrl@{}}
\caption{Discovery of the isotopes of all elements. The total number of isotopes, lightest and heaviest isotope and the year of first and most recent discovery is listed. The last column refers to the publication where the details of the discoveries are compiled.}\\
Element & Z & No.~of Iso. & Lightest & Heaviest & First & Last & Ref.  \\
\hline\\
\endfirsthead\\
\caption[]{(continued)}\\
Element & Z & No.~of Iso. & Lightest & Heaviest & First & Last & Ref.  \\
\hline\\
\endhead
Neutron(s) 	&	0	&	2	&	1	&	2	&	1932	&	1965	&	 \cite{2012Tho01} \\
Hydrogen 	&	1	&	7	&	1	&	7	&	1920	&	2003	&	 \cite{2012Tho01} \\
Helium 	&	2	&	9	&	2	&	10	&	1908	&	1994	&	 \cite{2012Tho01} \\
Lithium 	&	3	&	10	&	4	&	13	&	1921	&	2008	&	 \cite{2012Tho01} \\
Beryllium 	&	4	&	9	&	6	&	14	&	1921	&	1983	&	 \cite{2012Tho01} \\
Boron 	&	5	&	13	&	7	&	19	&	1920	&	2010	&	 \cite{2012Tho01} \\
Carbon 	&	6	&	14	&	8	 &	22	&	1919	&	1986	&	 \cite{2012Tho01} \\
Nitrogen 	&	7	&	14	&	10	&	23	&	1920	&	2002	&	 \cite{2012Tho01} \\
Oxygen 	&	8	&	14	&	12	&	25	&	1919	&	2008	&	 \cite{2012Tho01} \\
Fluorine 	&	9	&	16	&	14	&	31	&	1920	&	2010	&	 \cite{2012Tho01} \\
Neon 	&	10	&	18	&	16	&	34	&	1913	&	2002	&	 \cite{2012Tho01} \\
Sodium 	&	11	&	19	&	18	&	37	&	1921	&	2004	&	 \cite{2012Tho02} \\
Magnesium 	&	12	&	21	&	19	&	40	&	1920	&	2007	&	 \cite{2012Tho02} \\
Aluminum 	&	13	&	22	&	22	&	43	&	1922	&	2007	&	 \cite{2012Tho02} \\
Silicon 	&	14	&	23	&	22	&	44	&	1920	&	2007	&	 \cite{2012Tho02} \\
Phosphorus 	&	15	&	21	&	26	&	46	&	1920	&	1990	&	 \cite{2012Tho02} \\
Sulfur 	&	16	&	22	&	27	&	48	&	1920	&	1990	&	 \cite{2012Tho02} \\
Chlorine 	&	17	&	21	&	31	&	51	&	1919	&	2009	&	 \cite{2012Tho02} \\
Argon 	&	18	&	23	&	31	&	53	&	1920	&	2009	&	 \cite{2012Tho02} \\
Potassium 	&	19	&	22	&	35	&	56	&	1921	&	2009	&	 \cite{2012Tho02} \\
Calcium 	&	20	&	24	&	35	&	58	&	1922	&	2009	&	 \cite{2011Amo01} \\
Scandium 	&	21	&	23	&	39	&	61	&	1923	&	2009	&	 \cite{2011Mei01} \\
Titanium 	&	22	&	25	&	39	&	63	&	1923	&	2009	&	 \cite{2011Mei01} \\
Vanadium 	&	23	&	24	&	43	&	66	&	1923	&	2009	&	 \cite{2010Sho02} \\
Chromium 	&	24	&	27	&	42	&	68	&	1923	&	2009	&	 \cite{2012Gar01} \\
Manganese 	&	25	&	26	&	46	&	71	&	1923	&	2010	&	 \cite{2012Gar01} \\
Iron 	&	26	&	30	&	45	&	74	&	1922	&	2010	&	 \cite{2010Sch01} \\
Cobalt 	&	27	&	27	&	50	&	76	&	1923	&	2010	&	 \cite{2010Szy01} \\
Nickel 	&	28	&	32	&	48	&	79	&	1921	&	2010	&	 \cite{2012Gar01} \\
Copper 	&	29	&	28	&	55	&	82	&	1923	&	2010	&	 \cite{2012Gar01} \\
Zinc 	&	30	&	32	&	54	&	85	&	1922	&	2010	&	 \cite{2012Gro01} \\
Gallium 	&	31	&	28	&	60	&	87	&	1923	&	2010	&	 \cite{2012Gro02} \\
Germanium 	&	32	&	31	&	60	&	90	&	1923	&	2010	&	 \cite{2012Gro02} \\
Arsenic 	&	33	&	29	&	64	&	92	&	1920	&	1997	&	 \cite{2010Sho03} \\
Selenium 	&	34	&	32	&	64	&	95	&	1922	&	2010	&	 \cite{2012Gro01} \\
Bromine 	&	35	&	30	&	69	&	98	&	1920	&	2011	&	 \cite{2012Gro01} \\
Krypton 	&	36	&	33	&	69	&	101	&	1920	&	2010	&	 \cite{2010Hei01} \\
Rubidium 	&	37	&	31	&	73	&	103	&	1921	&	2010	&	 \cite{2012Par01} \\
Strontium 	&	38	&	35	&	73	&	107	&	1923	&	2010	&	 \cite{2012Par01} \\
Yttrium 	&	39	&	34	&	76	&	109	&	1923	&	2010	&	 \cite{2012Nys01} \\
Zirconium 	&	40	&	35	&	78	&	112	&	1924	&	2010	&	 \cite{2012Nys01} \\
Niobium 	&	41	&	34	&	82	&	115	&	1932	&	2010	&	 \cite{2012Nys01} \\
Molybdenum 	&	42	&	35	&	83	&	117	&	1930	&	2010	&	 \cite{2012Par01} \\
Technetium 	&	43	&	35	&	86	&	120	&	1938	&	2010	&	 \cite{2012Nys01} \\
Ruthenium 	&	44	&	38	&	87	&	124	&	1931	&	2010	&	 \cite{2012Nys01} \\
Rhodium 	&	45	&	38	&	89	&	126	&	1934	&	2010	&	 \cite{2012Par01} \\
Palladium 	&	46	&	38	&	91	&	128	&	1935	&	2010	&	 \cite{2013Kat01} \\
Silver 	&	47	&	38	&	93	&	130	&	1923	&	2000	&	 \cite{2010Sch03} \\
Cadmium 	&	48	&	38	&	96	&	133	&	1924	&	2010	&	 \cite{2010Amo01} \\
Indium 	&	49	&	38	&	98	&	135	&	1924	&	2002	&	 \cite{2011Amo01} \\
Tin 	&	50	&	39	&	100	&	138	&	1922	&	2010	&	 \cite{2011Amo01} \\
Antimony 	&	51	&	38	&	103	&	140	&	1922	&	2010	&	 \cite{2013Kat01} \\
Tellurium 	&	52	&	39	&	105	&	143	&	1924	&	2010	&	 \cite{2013Kat01} \\
Iodine 	&	53	&	38	&	108	&	145	&	1920	&	2010	&	 \cite{2013Kat01} \\
Xenon 	&	54	&	40	&	109	&	148	&	1920	&	2010	&	 \cite{2013Kat01} \\
Cesium 	&	55	&	41	&	112	&	152	&	1921	&	1994	&	 \cite{2012May01} \\
Barium 	&	56	&	39	&	114	&	152	&	1924	&	2010	&	 \cite{2010Sho01} \\
Lanthanum 	&	57	&	35	&	117	&	153	&	1924	&	2001	&	 \cite{2012May01} \\
Cerium 	&	58	&	35	&	121	&	155	&	1924	&	2005	&	 \cite{2009Gin01} \\
Praseodymium 	&	59	&	32	&	121	&	154	&	1924	&	2005	&	 \cite{2012May01} \\
Neodymium 	&	60	&	31	&	125	&	156	&	1924	&	1999	&	 \cite{2012Gro01} \\
Promethium 	&	61	&	32	&	128	&	159	&	1947	&	2005	&	 \cite{2012May01} \\
Samarium 	&	62	&	34	&	129	&	162	&	1933	&	2005	&	 \cite{2013May01} \\
Europium 	&	63	&	35	&	130	&	166	&	1933	&	2008	&	 \cite{2013May01} \\
Gadolinium 	&	64	&	31	&	135	&	166	&	1933	&	2005	&	 \cite{2013May01} \\
Terbium 	&	65	&	31	&	135	&	168	&	1933	&	2004	&	 \cite{2013May01} \\
Dysprosium 	&	66	&	32	&	139	&	170	&	1934	&	2010	&	 \cite{2013Fry01} \\
Holmium 	&	67	&	32	&	140	&	172	&	1934	&	2001	&	 \cite{2013Fry01} \\
Erbium 	&	68	&	32	&	144	&	175	&	1934	&	2003	&	 \cite{2013Fry01} \\
Thulium 	&	69	&	33	&	145	&	177	&	1934	&	1998	&	 \cite{2013Fry01} \\
Ytterbium 	&	70	&	31	&	149	&	180	&	1934	&	2001	&	 \cite{2013Fry01} \\
Lutetium 	&	71	&	35	&	150	&	184	&	1934	&	1993	&	 \cite{2012Gro02} \\
Hafnium 	&	72	&	36	&	154	&	189	&	1934	&	2009	&	 \cite{2012Gro02} \\
Tantalum 	&	73	&	38	&	155	&	192	&	1932	&	2009	&	 \cite{2012Rob01} \\
Tungsten 	&	74	&	38	&	157	&	194	&	1930	&	2010	&	 \cite{2010Fri01} \\
Rhenium 	&	75	&	39	&	159	&	197	&	1931	&	2011	&	 \cite{2012Rob01} \\
Osmium 	&	76	&	41	&	161	&	201	&	1931	&	2011	&	 \cite{2012Rob01} \\
Iridium 	&	77	&	40	&	165	&	204	&	1935	&	2011	&	 \cite{2012Rob01} \\
Platinum 	&	78	&	40	&	166	&	205	&	1935	&	2010	&	 \cite{2011Amo01} \\
Gold 	&	79	&	41	&	170	&	210	&	1935	&	2011	&	 \cite{2010Sch02} \\
Mercury 	&	80	&	46	&	171	&	216	&	1920	&	2010	&	 \cite{2011Mei01} \\
Thallium 	&	81	&	42	&	176	&	217	&	1908	&	2010	&	 \cite{2013Fry02} \\
Lead 	&	82	&	42	&	179	&	220	&	1900	&	2010	&	 \cite{2013Fry02} \\
Bismuth 	&	83	&	41	&	184	&	224	&	1904	&	2010	&	 \cite{2013Fry02} \\
Polonium 	&	84	&	42	&	186	&	227	&	1898	&	2010	&	 \cite{2013Fry02} \\
Astatine 	&	85	&	39	&	191	&	229	&	1940	&	2010	&	 \cite{2013Fry03} \\
Radon 	&	86	&	39	&	193	&	231	&	1899	&	2010	&	 \cite{2013Fry03} \\
Francium 	&	87	&	35	&	199	&	233	&	1939	&	2010	&	 \cite{2013Fry03} \\
Radium 	&	88	&	34	&	201	&	234	&	1898	&	2005	&	 \cite{2013Fry03} \\
Actinium 	&	89	&	31	&	206	&	236	&	1902	&	2010	&	 \cite{2013Fry04} \\
Thorium 	&	90	&	31	&	208	&	238	&	1898	&	2010	&	 \cite{2013Fry04} \\
Protactinium 	&	91	&	28	&	212	&	239	&	1913	&	2005	&	 \cite{2013Fry04} \\
Uranium 	&	92	&	23	&	217	&	242	&	1896	&	2000	&	 \cite{2013Fry04} \\
Neptunium 	&	93	&	20	&	225	&	244	&	1940	&	1994	&	 \cite{2013Fry05} \\
Plutonium 	&	94	&	20	&	228	&	247	&	1946	&	1999	&	 \cite{2013Fry05} \\
Americium 	&	95	&	16	&	232	&	247	&	1949	&	2000	&	 \cite{2013Fry05} \\
Curium 	&	96	&	17	&	233	&	251	&	1949	&	2010	&	 \cite{2013Fry05} \\
Berkelium 	&	97	&	13	&	238	&	251	&	1950	&	2003	&	 \cite{2013Fry05} \\
Californium 	&	98	&	20	&	237	&	256	&	1951	&	1995	&	 \cite{2013Fry05} \\
Einsteinium 	&	99	&	17	&	241	&	257	&	1954	&	1996	&	 \cite{2011Mei01} \\
Fermium 	&	100	&	19	&	241	&	259	&	1954	&	2008	&	 \cite{2013Tho01} \\
Mendelevium 	&	101	&	16	&	245	&	260	&	1955	&	1996	&	 \cite{2013Tho01} \\
Nobelium 	&	102	&	11	&	250	&	260	&	1963	&	2001	&	 \cite{2013Tho01} \\
Lawrencium 	&	103	&	9	&	252	&	260	&	1965	&	2001	&	 \cite{2013Tho01} \\
Rutherfordium 	&	104	&	13	&	253	&	267	&	1969	&	2010	&	 \cite{2013Tho01} \\
Dubnium 	&	105	&	11	&	256	&	270	&	1970	&	2010	&	 \cite{2013Tho01} \\
Seaborgium 	&	106	&	12	&	258	&	271	&	1974	&	2010	&	 \cite{2013Tho01} \\
Bohrium 	&	107	&	10	&	260	&	274	&	1981	&	2010	&	 \cite{2013Tho01} \\
Hassium 	&	108	&	12	&	263	&	277	&	1984	&	2010	&	 \cite{2013Tho01} \\
Meitnerium 	&	109	&	7	&	266	&	278	&	1982	&	2010	&	 \cite{2013Tho01} \\
Darmstadtium 	&	110	&	8	&	267	&	281	&	1995	&	2010	&	 \cite{2013Tho01} \\
Roengtenium 	&	111	&	7	&	272	&	282	&	1995	&	2010	&	 \cite{2013Tho01} \\
Copernicium 	&	112	&	6	&	277	&	285	&	1996	&	2010	&	 \cite{2013Tho01} \\
113	&	113	&	6	&	278	&	286	&	2004	&	2010	&	 \cite{2013Tho01} \\
Flerovium 	&	114	&	5	&	285	&	289	&	2004	&	2010	&	 \cite{2013Tho01} \\
115	&	115	&	4	&	287	&	290	&	2004	&	2010	&	 \cite{2013Tho01} \\
Livermorium 	&	116	&	4	&	290	&	293	&	2004	&	2004	&	 \cite{2013Tho01} \\
117	&	117	&	2	&	293	&	294	&	2010	&	2010	&	 \cite{2013Tho01} \\
118	&	118	&	1	&	294	&	294	&	2006	&	2006	&	 \cite{2013Tho01} \\ \hline

 \label{t:elements}
\end{longtable}
}

The element with the most isotopes (46) presently known is mercury, followed by thallium, lead and polonium with 42 each. The element with the fewest isotopes is element 118 where only one isotope (A = 294) is presently known. The heaviest nuclides are $^{294}$117 and $^{294}$118. However, it should be stressed that the observation of elements 117 and 118 has not been accepted by IUPAC.

\section{Potential Discoveries in the Near Future}
\label{sec:potential}

The 3015 nuclides presently reported in the published literature still probably constitute less than 50\% of all nuclides that potentially could be observed. In the following subsections nuclides which should be discovered in the near future are discussed.

\subsection{Proceedings and internal reports}
\label{subsec:proc}

Until the end of 2011 twenty-six nuclides had only been reported in conference proceedings or internal reports. Table \ref{t:nonref-pub} lists these nuclides along with the author, year, laboratory, conference or report and reference of the discovery. Most of them were reported at least ten years ago, so that it is unlikely that these results will be published in refereed journals in the future. Conference proceedings quite often contain preliminary results and it is conceivable that these results then do not hold up for a refereed journal.

A curious case is the reported discovery of $^{155,156}$Pr and $^{157,158}$Nd in the proceedings of RNB-3 in 1996 \cite{1996Cza01} where these nuclides were included as newly discovered in a figure of the chart of nuclides. The authors also stated: ``In this first experiment, 54 new isotopes were discovered, ranging from $^{86}_{32}$Ge to $^{158}_{60}$Nd'' \cite{1996Cza01}. However, in the original publication only 50 new isotopes were listed and there was no evidence for the observation of any praseodymium or neodymium isotopes \cite{1994Ber01}. A modified version of the nuclide chart showing these nuclei was included in two further publications \cite{1997Ber01,1997Ber02}.

These two neodymium isotopes ($^{157,158}$Nd) have recently been reported  (see Section \ref{sec:2012}) by van Schelt et al. \cite{2012Van01} and Kurcewicz et al. \cite{2012Kur01}, respectively.

{\setlength{\LTleft}{0pt}
\setlength{\LTright}{0pt}


\setlength{\tabcolsep}{0.5\tabcolsep}

\renewcommand{\arraystretch}{1.0}
\renewcommand{\baselinestretch}{1.0}

\footnotesize 
\begin{longtable}{@{\extracolsep\fill}llllll@{}}
\caption{Nuclides only reported in proceedings or internal reports until the end of 2011. The nuclide, author, year, laboratory, conference or report and reference of the discovery are listed. }\\
Nuclide(s) & Author & Year & Laboratory & Conf./Report & Ref.\\
\hline\\
\endfirsthead\\
\caption[]{(continued)}\\
Nuclide(s) & Author & Year & Laboratory & Conf./Report & Ref.\\
\hline\\
\endhead
\renewcommand{\arraystretch}{1.9}
$^{95}$Cd,$^{97}$In\footnotetext{Discovered in 2012, see discussion in Section \ref{sec:2012}}$^{\footnotemark[1]}$  &  R. Kr\"ucken  & 2008 &  GSI  &  \parbox[t]{5cm}{\raggedright Nucl. Phys. and Astrophys.: From Stable Beams to Exotic Nuclei, 25-30 June 2008, Cappadocia (Turkey) \vspace*{0.2cm} }  &  \cite{2008Kru01} \\
$^{155}$Pr$^{\footnotemark[1]}$  $^{156}$Pr&  S. Czajkowski et al.  & 1996 &  GSI  &  \parbox[t]{5cm}{\raggedright ENAM'95, 19-23 June 1995, Arles (France) \vspace*{0.2cm}}  &  \cite{1996Cza01} \\
$^{126}$Nd  &  G. A. Souliotis  & 2000 &  MSU  &  \parbox[t]{5cm}{\raggedright Int. Conf. on Achievements and Perspectives in Nuclear Structure, 11-17 July 1999, Aghia Palaghia, Crete (Greece) \vspace*{0.2cm}}  &  \cite{2000Sou01} \\
$^{157,158}$Nd$^{\footnotemark[1]}$  &  S. Czajkowski et al.  & 1996 &  GSI  &  \parbox[t]{5cm}{\raggedright ENAM'95, 19-23 June 1995, Arles (France) \vspace*{0.2cm}}  &  \cite{1996Cza01} \\
$^{136}$Gd,$^{138}$Tb &  G. A. Souliotis  & 2000 &  MSU  &  \parbox[t]{5cm}{\raggedright Int. Conf. on Achievements and Perspectives in Nuclear Structure, 11-17 July 1999, Aghia Palaghia, Crete (Greece) \vspace*{0.2cm}}  &  \cite{2000Sou01} \\
$^{143}$Ho  &  G. A. Souliotis  & 2000 &  MSU  &  \parbox[t]{5cm}{\raggedright Int. Conf. on Achievements and Perspectives in Nuclear Structure, 11-17 July 1999, Aghia Palaghia, Crete (Greece) \vspace*{0.2cm}}  &  \cite{2000Sou01} \\
  &  D. Seweryniak et al.  & 2002 &  LBL  &  \parbox[t]{5cm}{\raggedright Annual Report \vspace*{0.2cm}}  &  \cite{2003Sew02} \\
$^{144}$Tm  &  K. P. Rykaczewski et al.  & 2004 &  ORNL  &  \parbox[t]{5cm}{\raggedright Nuclei at the Limits, 26-30 July 2004, Argonne, Illinois (USA) \vspace*{0.05cm} }  &  \cite{2005Ryk01} \\
 &  R. Grzywacz et al.  &   &   &  \parbox[t]{5cm}{\raggedright ENAM2004, September 12-16 2004, Pine Mountain, Georgia \vspace*{0.05cm}}  &  \cite{2005Grz01} \\
 &  C. R. Bingham et al.  &   &   &  \parbox[t]{5cm}{\raggedright CAARI2004, 10-15 October 2004, Fort Worth, Texas (USA) \vspace*{0.2cm}}  &  \cite{2005Bin01} \\
$^{178}$Tm$^{\footnotemark[1]}$  &  Zs. Podolyak et al.  & 1999 &  GSI  &  \parbox[t]{5cm}{\raggedright 2$^{nd}$ Int. Conf. Fission and Properties of Neutron-Rich Nuclei, June 28-July 3, 1999, St. Andrews (Scotland) \vspace*{0.2cm}}  &  \cite{2000Pod01} \\
$^{150}$Yb  &  G. A. Souliotis  & 2000 &  MSU  &  \parbox[t]{5cm}{\raggedright Int. Conf. on Achievements and Perspectives in Nuclear Structure, 11-17 July 1999, Aghia Palaghia, Crete (Greece) \vspace*{0.2cm}}  &  \cite{2000Sou01} \\
$^{181}$Yb$^{\footnotemark[1]}$  &  Zs. Podolyak et al.  & 1999 &  GSI  &  \parbox[t]{5cm}{\raggedright 2$^{nd}$ Int. Conf. Fission and Properties of Neutron-Rich Nuclei, June 28-July 3, 1999, St. Andrews (Scotland) \vspace*{0.2cm}}  &  \cite{2000Pod01} \\
$^{182}$Yb$^{\footnotemark[1]}$  &  S. D. Al-Garni et al.  & 2002 &  GSI  &  \parbox[t]{5cm}{\raggedright Annual Report \vspace*{0.2cm}}  &  \cite{2002AlG01} \\
$^{153}$Hf  &  G. A. Souliotis  & 2000 &  MSU  &  \parbox[t]{5cm}{\raggedright Int. Conf. on Achievements and Perspectives in Nuclear Structure, 11-17 July 1999, Aghia Palaghia, Crete (Greece) \vspace*{0.2cm}}  &  \cite{2000Sou01} \\
$^{164}$Ir  &  H. Kettunen et al.  & 2000 &  Jyv\"askyl\"a  &  \parbox[t]{5cm}{\raggedright XXXV Zakopane School of Physics, 5-13 September 2000, Zakopane (Poland) \vspace*{0.2cm}}  &  \cite{2001Ket02} \\
 &  H. Mahmud et al.  & 2001 &  ANL  &  \parbox[t]{5cm}{\raggedright ENAM2001, 2-7 July 2001, H\"ameenlinna (Finland) \vspace*{0.05cm} }  &  \cite{2002Mah01} \\
 &  D. Seweryniak et al.  &   &   &  \parbox[t]{5cm}{\raggedright Frontiers of Nuclear Structure, 29 July - 2 August 2002, Berkeley, California (USA) \vspace*{0.2cm}}  &  \cite{2003Sew01} \\
$^{234}$Cm  &  P. Cardaja et al.  & 2002 &  GSI  &  \parbox[t]{5cm}{\raggedright Annual Report \vspace*{0.2cm}}  &  \cite{2002Cag01} \\
 &  J. Khuyagbaatar et al.  & 2007 &  GSI  &  \parbox[t]{5cm}{\raggedright Annual Report \vspace*{0.2cm}}  &  \cite{2007Khu01} \\
 &  D. Kaji et al.  & 2010 &  RIKEN  &  \parbox[t]{5cm}{\raggedright Annual Report \vspace*{0.2cm}}  &  \cite{2010Kaj01} \\
$^{235}$Cm  &  J. Khuyagbaatar et al.  & 2007 &  GSI  &  \parbox[t]{5cm}{\raggedright Annual Report \vspace*{0.2cm}}  &  \cite{2007Khu01} \\
$^{234}$Bk  &  K. Morita et al.  & 2002 &  RIKEN  &  \parbox[t]{5cm}{\raggedright Frontiers of Collective Motion, 6-9 November 2002, Aizu (Japan) \vspace*{0.05cm}}  &  \cite{2003Mor02} \\
 &  K. Morimoto et al.  &   &   &  \parbox[t]{5cm}{\raggedright Annual Report \vspace*{0.2cm}}  &  \cite{2003Mor01} \\
 &  D. Kaji et al.  & 2010 &  RIKEN  &  \parbox[t]{5cm}{\raggedright Annual Report \vspace*{0.2cm}}  &  \cite{2010Kaj01} \\
$^{252,253}$Bk  &  S. A. Kreek et al.  & 1992 &  LBL  &  \parbox[t]{5cm}{\raggedright Annual Report }  &  \cite{1992Kre01} \\
$^{262}$No  &  R. W. Lougheed et al.  & 1988 &  LBL  &  \parbox[t]{5cm}{\raggedright Annual Report }  &  \cite{1988Lou01} \\
 &   &   &   &  \parbox[t]{5cm}{\raggedright 50 years with nuclear fission, April 25-28, 1989, Gaithersburg, Maryland (USA) \vspace*{0.05cm}}  &  \cite{1989Lou01} \\
 &  E. K. Hulet  &   &   &  \parbox[t]{5cm}{\raggedright Internal Report \vspace*{0.2cm} }  &  \cite{1989Hul01} \\
$^{261}$Lr  &  R. W. Lougheed et al.  & 1987 &  LBL  &  \parbox[t]{5cm}{\raggedright Annual Report \vspace*{0.05cm}}  &  \cite{1987Lou01} \\
 &  E. K. Hulet  &   &   &  \parbox[t]{5cm}{\raggedright Internal Report \vspace*{0.2cm}}  &  \cite{1989Hul01} \\
 &  R. A. Henderson et al.  & 1991 &  LBL  &  \parbox[t]{5cm}{\raggedright Annual Report \vspace*{0.2cm}}  &  \cite{1991Hen01} \\
$^{262}$Lr  &  R. W. Lougheed et al.  & 1987 &  LBL  &  \parbox[t]{5cm}{\raggedright Annual Report \vspace*{0.05cm}}  &  \cite{1987Lou01} \\
 &  E. K. Hulet  &   &   &  \parbox[t]{5cm}{\raggedright Internal Report \vspace*{0.2cm}}  &  \cite{1989Hul01} \\
 &  R. A. Henderson et al.  & 1991 &  LBL  &  \parbox[t]{5cm}{\raggedright Annual Report \vspace*{0.2cm}}  &  \cite{1991Hen01} \\
$^{255}$Db  &  G. N. Flerov  & 1976 &  Dubna  &  \parbox[t]{5cm}{\raggedright 3$^{rd}$ Int. Conf. on Nuclei Far from Stability, 19-26 May 1976, Cargese, Corsica (France) \vspace*{0.2cm}}  &  \cite{1976Fle01} \\ \hline
\\
\label{t:nonref-pub}
\end{longtable}
}

Another argument for not giving full credit for a discovery reported in conference proceedings are contributions from single authors (for example \cite{2008Kru01,2000Sou01}). These experiments typically involve fairly large collaborations and it is not clear that these single-author papers were fully vetted by the collaboration. Also everyone involved in the experiment and the analysis should get the appropriate credit.

The authors of the more recent proceedings and reports are encouraged to fully analyze the data and submit their final results for publication in refereed journals.

\subsection{Medium-mass proton rich nuclides}
\label{subsec:proton}

The proton dripline has been crossed in the medium-mass region between antimony and bismuth (Z = 51--83) with the observation of proton emitters of odd-Z elements. Promethium is the only odd-Z element in this mass region where no proton emitters have been discovered yet. In these experiments the protons are detected in position sensitive silicon detectors correlated with the implantation of a fusion-evaporation residue after a mass separator. The high detection efficiency for these protons makes this method very efficient and nuclides far beyond the proton dripline with very small cross sections can be identified.

In contrast, for nuclides closer to the dripline proton emission is not the dominant decay mode due to the smaller Q-values for the proton decay. The identification of these nuclides is more difficult because of the lower detection efficiency for $\beta$- and $\gamma$-rays. In fact many of these nuclei were identified by $\beta$-delayed proton emission from excited states of the daughter nuclei. Thus, there are isotopes not yet discovered between the lightest $\beta$-emitters and the heaviest proton emitters for the odd-Z elements.

\begin{figure}
	\centering
	\includegraphics[scale=.6]{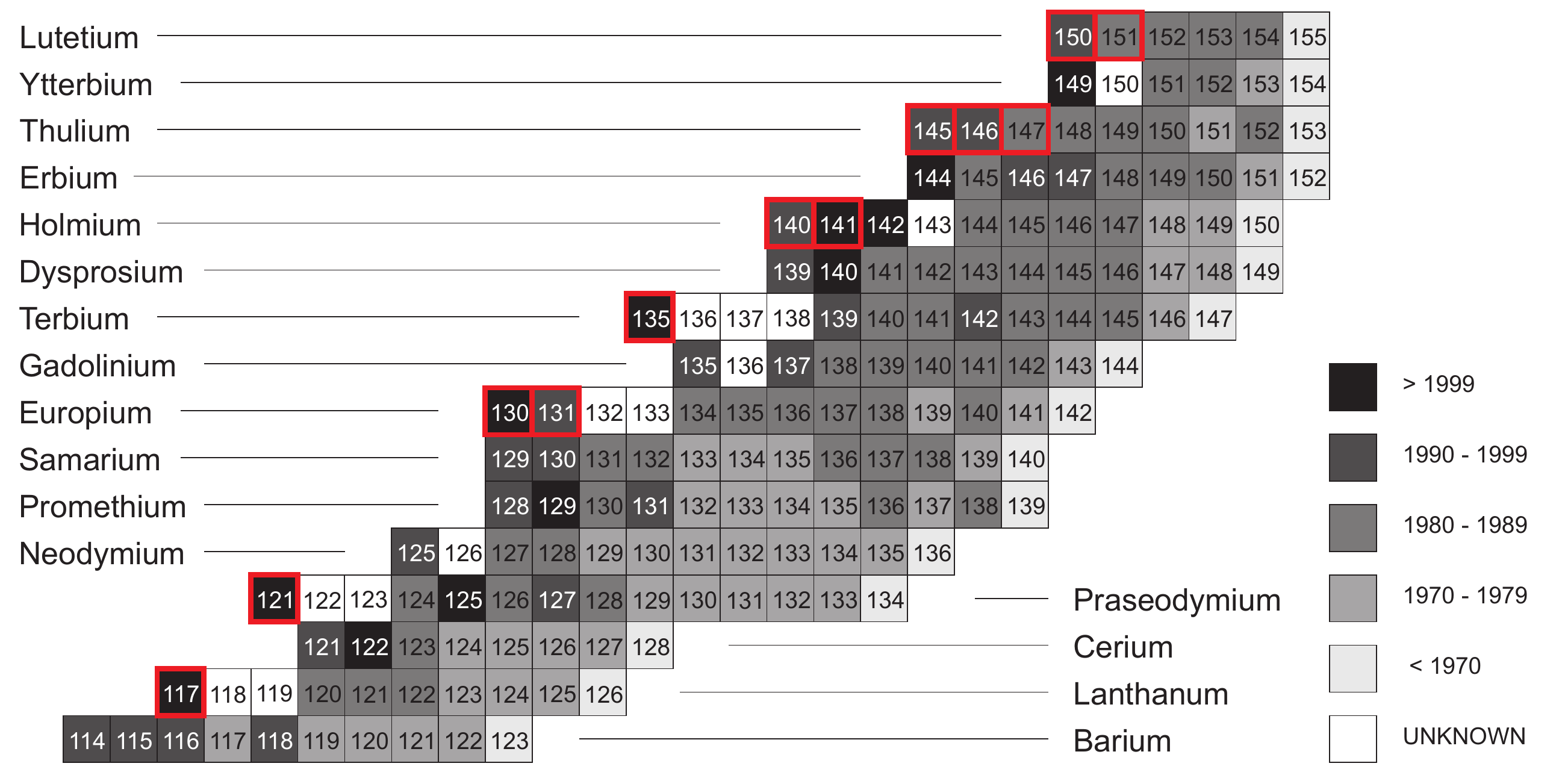}
	\caption{Chart of nuclides for neutron-deficient nuclides between barium and lutetium (Z = 56--71. The grey-scale coding refers to the decade of discovery. Proton emitters are identified by the thick (red) borders.}
\label{f:proton-holes}
\end{figure}

Figure \ref{f:proton-holes} shows the medium-mass neutron-deficient region of the chart of nuclides. The thick (red) borders indicate proton emitters and the grey shades of the nuclides indicate the decade of discovery.

Currently nine odd-Z nuclides ($^{118,119}$La, $^{122,123}$La, $^{132,133}$Eu, and $^{136,137,138}$Tb) fall into these gaps. For the tenth missing nuclide, $^{143}$Ho, $^{142}$Ho has already been identified by $\beta$-delayed proton emission \cite{2001Xu02}. In fact, decay properties of $^{143}$Ho have also been measured but the results were only reported in an annual report \cite{2003Sew02}.

There are three even-Z holes in this mass region: $^{126}$Nd, $^{136}$Gd, and $^{150}$Yb.  In all three cases, the even more neutron-deficient nuclides were observed by the detection of $\beta$-delayed proton emission at the Institute of Modern Physics, Lanzhou, China ($^{125}$Nd \cite{1999Xu01}, $^{135}$Gd \cite{1996Xu01}, and $^{149}$Yb \cite{2001Xu01}).

The identification of $^{126}$Nd, $^{136}$Gd, and $^{150}$Yb in the fragmentation reaction of a 30 MeV/nucleon $^{197}$Au beam has been reported only in a contribution to a conference proceeding \cite{2000Sou01}. The recent advances in beam intensities and detection techniques for fragmentation reactions (especially identification and separation of charge states) should make it possible to discover these and many more additional nuclides along and beyond the proton dripline in this mass region.

\subsection{Medium mass neutron rich nuclei}
\label{subsec:neutron}

In contrast to the proton dripline the neutron dripline has not been reached for medium mass nuclides. The heaviest neutron-rich nuclide shown to be unbound is $^{39}$Mg \cite{2002Not01}. Most of the most neutron-rich nuclides have been produced in projectile fragmentation or projectile fission over the last fifteen years. The nuclides are separated with fragment separators according to their magnetic rigidity ( = momentum over charge of the nuclides which corresponds approximately to their A/Z) and identified by time-of-flight and energy-loss measurements.

\begin{figure}
	\centering
	\includegraphics[scale=.9]{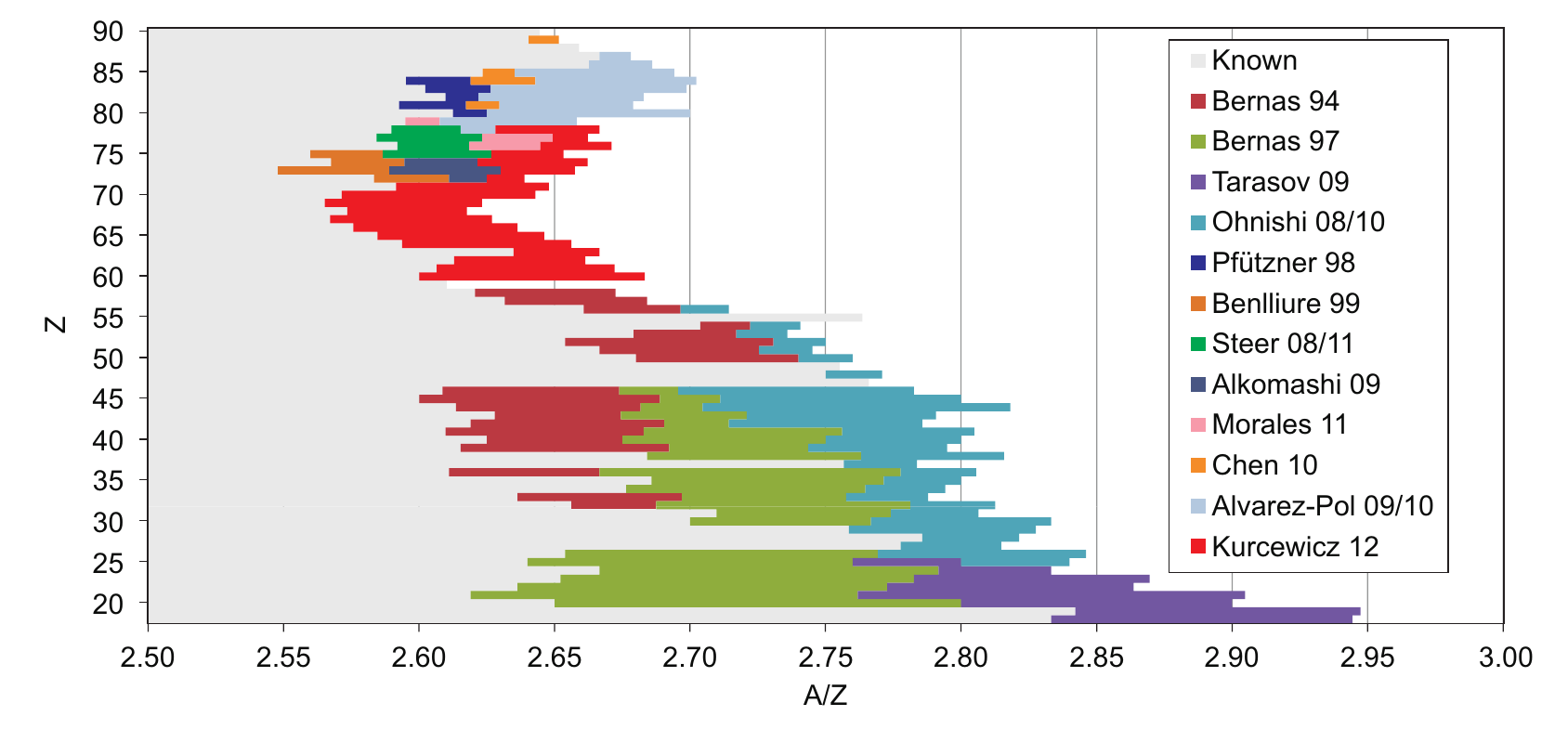}
	\caption{Neutron-rich nuclei between argon and thorium discovered by projectile fragmentation or projectile fission as a function of A/Z. The data as labeled in the figure are from Bernas 94 \cite{1994Ber01}, Bernas 97 \cite{1997Ber01}, Tarasov 09 \cite{2009Tar01}, Ohnishi 08/10 \cite{2008Ohn01,2010Ohn01}, Pf\"utzner 98 \cite{1998Pfu01}, Benlliure 99 \cite{1999Ben01}, Steer 08/11\cite{2008Ste01,2011Ste01}, Alkomashi 09 \cite{2009Alk01}, Morales 11 \cite{2011Mor01}, Chen 10 \cite{2010Che01}, Alvarez-Pol 09/10 \cite{2009Alv01,2010Alv01}, and Kurcewicz \cite{2012Kur01}. The legend in the figure refers to the first author and year of the publications.}
\label{f:brho}
\end{figure}

Figure \ref{f:brho} displays the neutron-rich part of the chart of nuclides between argon and thorium (Z = 18--90) as a function of A/Z. It shows the A/Z ranges covered by the different experiments. The figure also includes the most recent measurement by Kurcewicz et al. \cite{2012Kur01} (see Section \ref{sec:2012}). If one considers that the location of the neutron dripline is predicted to be more or less constant at about 3.2 in this mass region, it is clear from the figure that it is still far away. The limits of the projectile fragmentation/fission method is presently determined by the small cross sections which can be overcome to an extend by improvements of primary beam intensities and/or larger acceptance separators. In the long term the method is limited by the limited availability of neutron rich projectiles.

\subsection{Superheavy nuclides}
\label{subsec:heavy}

\begin{figure}[tbh]
	\centering
	\includegraphics[scale=.6]{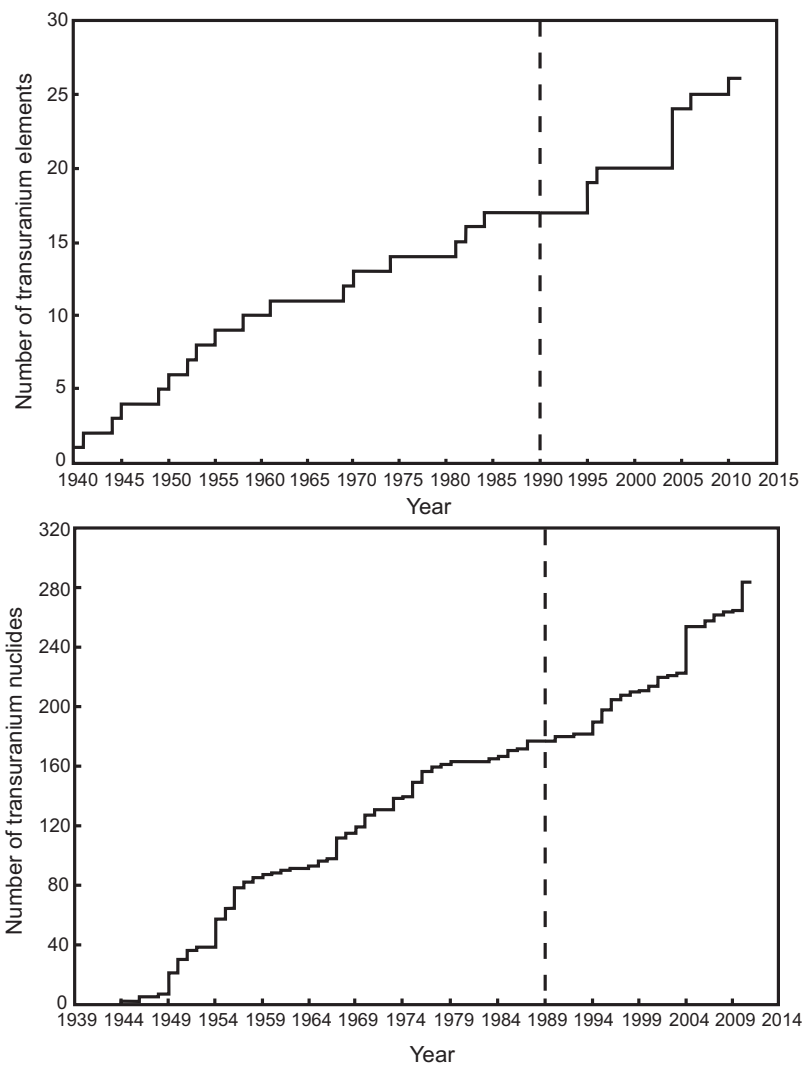}
	\caption{Number of discovered transuranium elements and nuclides. The data until 1989/1990 indicated by the dashed line were taken from Ref. \cite{1990Sea01}.}
\label{f:seaborg-elements}
\end{figure}

The discovery of superheavy nuclides has always been special because it is directly related to the discovery of new elements. It is interesting to follow the evolution of element discovery and the discovery of nuclides. In the 1990 book ``The elements beyond uranium'' Seaborg and Loveland showed the number of discovered transuranium elements and nuclides as a function of year \cite{1990Sea01}. Figure \ref{f:seaborg-elements} displays an extention of these data until today. The number of discovered nuclides tracks closely the number of discovered elements with about 10 isotopes per elements.

\begin{figure}
	\centering
	\includegraphics[scale=.6]{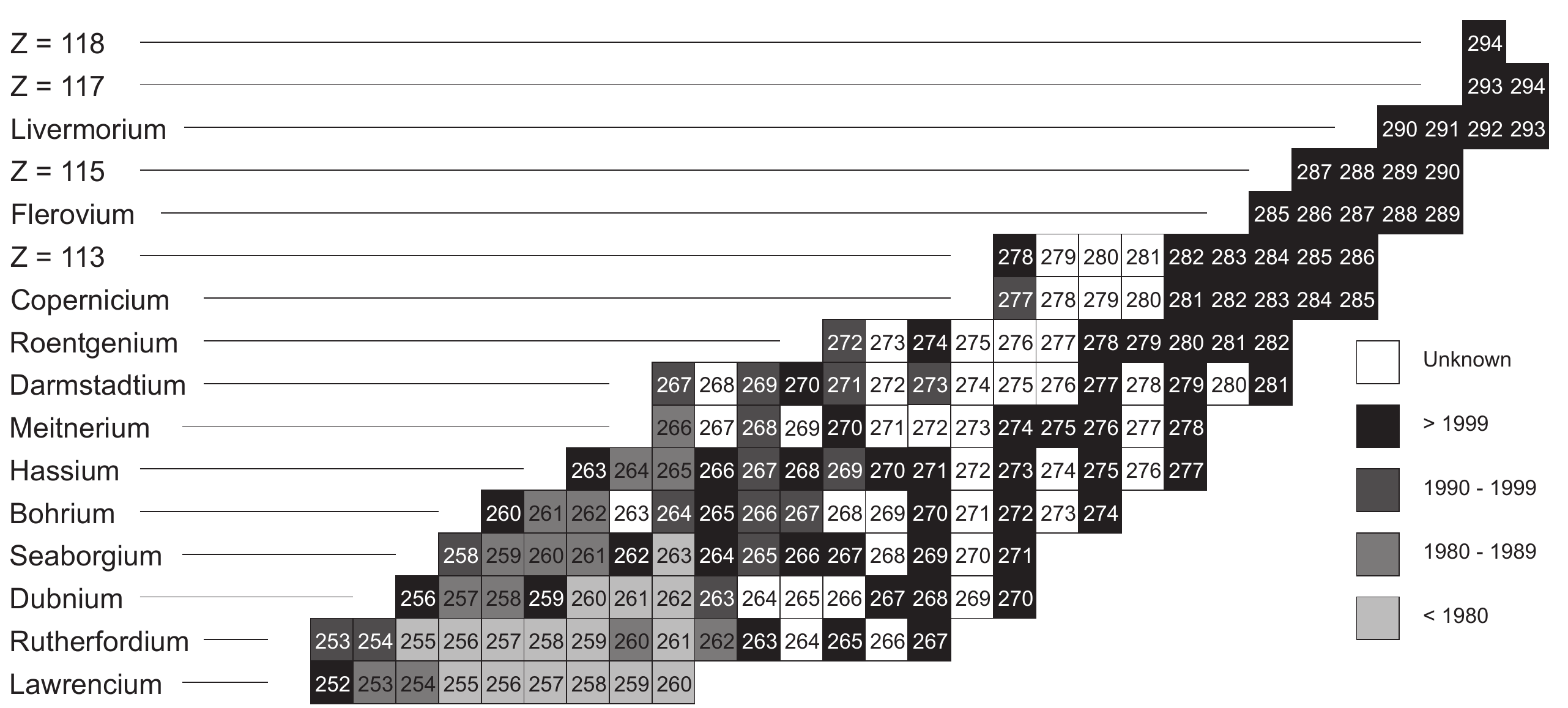}
	\caption{Chart of nuclides for elements heavier than nobelium. The grey-scale coding refers to the decade of discovery.}
\label{f:superheavy-gaps}
\end{figure}

In addition to the efforts to discover elements 119 and 120 it is important to link the isotopes of the elements beyond 113 to known nuclides. Figure \ref{f:superheavy-gaps} shows the nuclear chart beyond nobelium. It shows the separation of the more neutron-rich nuclides up to Z = 118 produced in ``hot'' fusion evaporation reactions from the less neutron-rich nuclides up to Z = 113 which were predominantly produced in ``cold'' fusion evaporation reactions. No isotone with N = 164 has been observed so far which does not mean that this isotone line corresponds to the separation of the decay chains.

Table \ref{t:chains} lists the 10 presently observed unconnected decay chains. There are five even-Z and five odd-Z chains. The four chains starting at $^{282}$113, $^{285}$Fl, $^{288}$115, and $^{291}$Fl bridge the N = 164 gap and end in $^{270}$Bh, $^{265}$Rf, $^{268}$Db, and $^{267}$Rf, respectively. It should be mentioned that the odd-Z N--Z = 57 chain passes through the N = 164 isotone $^{271}$Bh, however, the properties of this nuclide could not unambiguously be determined \cite{2013Tho01}.

\begin{table}
\caption{Unconnected superheavy decay chains. The (N--Z) value, first and last nuclides and the number of $\alpha$-decays in the chains are listed.}
\begin{center}
\begin{tabular}{lcllc}
 & (N--Z) Chain & First & Last &  of $\alpha$ decays \\ \hline
Even-Z & 57 & $^{285}$Fl & $^{265}$Rf & 5 \\
 & 58 & $^{294}$118 & $^{282}$Cn & 3 \\
 & 59 & $^{291}$Lv & $^{267}$Rf & 6 \\
 & 60 & $^{292}$Lv & $^{284}$Cn & 2 \\
 & 61 & $^{293}$Lv & $^{277}$Hs & 4 \\ \hline
Odd-Z & 56 & $^{282}$113 & $^{266}$Db & 4 \\
 & 57 & $^{287}$115 & $^{267}$Db & 5 \\
 & 58 & $^{288}$115 & $^{268}$Db & 5 \\
 & 59 & $^{293}$117 & $^{281}$Rg & 3 \\
 & 60 & $^{294}$117& $^{270}$Db  & 6 \\ \hline
\end{tabular}
\end{center}
\label{t:chains}
\end{table}

The decay chains cannot be connected to known nuclides by extending them to lower masses because they terminate in nuclides which spontaneously fission. The relationship has to be established by systematic features of neighboring isotopes for different elements. Thus the missing isotopes $^{279-281}$113, $^{278-280}$Cn, $^{275-277}$Rg, and $^{274-276}$Ds as well as the other N = 164 isotopes $^{273}$Mt, $^{272}$Hs, $^{271}$Bh, $^{270}$Sg, and $^{269}$Db have to measured. In total there are thirtynine nuclides still to be discovered between already known light and heavy rutherfordium and Z = 113 nuclides.

In addition, there are a few gaps of unknown nuclides in the lighter (trans)uranium region. $^{239}$Bk and the two curium isotopes $^{234,235}$Cm have yet to be discovered, although as mentioned in Section \ref{subsec:proc} the curium isotopes have been reported in annual reports. Also three uranium isotopes are still unknown, $^{220,221}$U and $^{241}$U. It is especially surprising that the two lighter isotopes $^{220,221}$U have not been observed because three even lighter isotopes ($^{217-219}$U) are known. $^{222}$U was formed in the fusion evaporation reaction $^{186}$W($^{40}$Ar,4n) \cite{1983Hin01} and most of the other light uranium isotopes were formed in 4n or 5n reactions. Thus  $^{220,221}$U should be able to be populated and identified in $^{184}$W($^{40}$Ar,4n) and $^{186}$W($^{40}$Ar,5n) reactions, respectively.

\subsection{Beyond the driplines}
\label{subsec:limits}

As mentioned in Section \ref{sec:status} the present definition of nuclides also includes very short-lived nuclides beyond the proton- and neutron driplines. So far, these nuclides are only accessible in the light mass region and characteristics of many of these nuclides up to magnesium beyond the proton dripline and up to oxygen beyond the neutron dripline have been measured.

The proton dripline has most likely been reached or crossed for all elements up to technetium (Z = 43). Table \ref{t:proton-unbound} lists the first isotope of elements between aluminum and technetium which has been shown to be unbound but which has not been identified yet or the first isotope for which nothing is known, so that in principle it still could be bound or could have a finite lifetime. With maybe the exception of scandium, bromine, and rubidium where resonances have been already measured for $^{38}$Sc \cite{1988Woo01}, $^{69}$Br \cite{2011Rog01} and $^{73}$Rb \cite{1993Bat02}, resonance parameters for at least one isotope of these elements should be in reach in the near future.

For elements lighter than aluminum at least one unbound isotope has been identified. Although not impossible it is unlikely that further nuclides will exist for which characteristic resonance parameters can be measured.

{\setlength{\LTleft}{0pt}
\setlength{\LTright}{0pt}


\setlength{\tabcolsep}{0.5\tabcolsep}

\renewcommand{\arraystretch}{1.0}
\renewcommand{\baselinestretch}{1.0}

\footnotesize 
\begin{longtable}{@{\extracolsep\fill}lllllll@{}}

\caption{Nuclides beyond the proton dripline which have been demonstrated to be unbound or have not been reported yet.} \\
Z & Nuclide & Author & Year & Laboratory & Comments & Ref.\\
\hline\\
\endfirsthead\\
\caption[]{(continued)} \\
Z & Nuclide & Author & Year & Laboratory & Comments & Ref.\\
\hline\\
\endhead
13	 & 	$^{21}$Al	 & 		 & 		 & 		 & 	not measured	 & 		  		 \\
14	 & 	$^{21}$Si	 & 		 & 		 & 		 & 	not measured	 & 		 		 \\
15	 & 	$^{25}$P	 & 	M. Langevin	 & 	1986	 & 	GANIL	 & 		 & 	\cite{1986Lan01}	 	 \\
16	 & 	$^{26}$S	 & 	A.S. Fomichev	 & 	2011	 & 	Dubna	 & 		 & 	\cite{2011Fom01}	 	 \\
17	 & 	$^{29}$Cl	 & 	M. Langevin	 & 	1986	 & 	GANIL	 & 		 & 	\cite{1986Lan01}		 \\
	 & 	$^{30}$Cl	 & 	M. Langevin	 & 	1986	 & 	GANIL	 & 		 & 	\cite{1986Lan01}		 \\
18	 & 	$^{30}$Ar	 & 		 & 		 & 		 & 	not measured	 & 		 	 \\
19	 & 	$^{33}$K	 & 	M. Langevin	 & 	1986	 & 	GANIL	 & 		 & 	\cite{1986Lan01}	 	 \\
	 & 	$^{34}$K	 & 	M. Langevin	 & 	1986	 & 	GANIL	 & 		 & 	\cite{1986Lan01}	 	 \\
20	 & 	$^{34}$Ca	 & 		 & 		 & 		 & 	not measured	 & 		 		 \\
21	 & 	$^{38}$Sc	 & 		 & 		 & 		 & 	not measured, but $^{39}$Sc unbound	 & 		 		 \\
22	 & 	$^{38}$Ti	 & 	B. Blank	 & 	1996	 & 	GANIL	 & 		 & 	\cite{1996Bla01}		 \\
23	 & 	$^{42}$V	 & 	V. Borrel	 & 	1992	 & 	GANIL	 & 		 & 	\cite{1992Bor01}	 	 \\
24	 & 	$^{41}$Cr	 & 		 & 		 & 		 & 	not measured	 & 		  		 \\
25	 & 	$^{44}$Mn	 & 	V. Borrel	 & 	1992	 & 	GANIL	 & 		 & 	\cite{1992Bor01}	 	 \\
	 & 	$^{45}$Mn	 & 	V. Borrel	 & 	1992	 & 	GANIL	 & 		 & 	\cite{1992Bor01}		 \\
26	 & 	$^{44}$Fe	 & 		 & 		 & 		 & 	not measured, but $^{45}$Fe 2p emitter	 & 		 		 \\
27	 & 	$^{49}$Co	 & 	B. Blank	 & 	1994	 & 	GANIL	 & 		 & 	\cite{1994Bla01}	 	 \\
28	 & 	$^{47}$Ni	 & 		 & 		 & 		 & 	not measured, but $^{48}$Ni 2p emitter	 & 		  		 \\
29	 & 	$^{54}$Cu	 & 	B. Blank	 & 	1994	 & 	GANIL	 & 		 & 	\cite{1994Bla01}		 \\
30	 & 	$^{53}$Zn	 & 		 & 		 & 		 & 	not measured, but $^{54}$Zn 2p emitter	 & 		 		 \\
31	 & 	$^{59}$Ga	 & 	A. Stolz	 & 	2005	 & 	MSU	 & 		 & 	\cite{2005Sto01}	 	 \\
32	 & 	$^{59}$Ge	 & 		 & 		 & 		 & 	not measured	 & 		 	 \\
33	 & 	$^{63}$As	 & 	A. Stolz	 & 	2005	 & 	MSU	 & 		 & 	\cite{2005Sto01}	 	 \\
34	 & 	$^{63}$Se	 & 		 & 		 & 		 & 	not measured	 & 			 \\
35	 & 	$^{68}$Br	 & 		 & 		 & 		 & 	not measured, but $^{69}$Br unbound	 & 			 \\
36	 & 	$^{68}$Kr	 & 		 & 		 & 		 & 	not measured	 & 				 \\
37	 & 	$^{72}$Rb	 & 		 & 		 & 		 & 	not measured, but $^{73}$Rb unbound	 & 				 \\
38	 & 	$^{72}$Sr	 & 		 & 		 & 		 & 	not measured	 & 				 \\
39	 & 	$^{75}$Y	 & 		 & 		 & 		 & 	not measured	 & 		 		 \\
40	 & 	$^{77}$Zr	 & 		 & 		 & 		 & 	not measured	 & 			 \\
41	 & 	$^{81}$Nb	 & 	Z. Janas	 & 	1999	 & 	GANIL	 & 		 & 	\cite{1999Jan01}	 \\
42	 & 	$^{82}$Mo	 & 		 & 		 & 		 & 	not measured	 & 		  \\
43	 & 	$^{85}$Tc	 & 	Z. Janas	 & 	1999	 & 	GANIL	 & 		 & 	\cite{1999Jan01}		 \\ \hline
\label{t:proton-unbound}
\end{longtable}
}

For neutron rich nuclei characteristic properties of at least two isotopes beyond the neutron dripline have been identified for the lightest elements, hydrogen, helium and lithium. Neutron rich nuclides between beryllium and magnesium which have been shown or expected to be unbound but have not been observed are listed in Table \ref{t:neutron-unbound}. Most of these nuclides should be able to be measured in the near future. Indeed, $^{16}$Be, $^{26}$O, and $^{28}$F have been discovered recently (see Section \ref{sec:2012}). The open question whether the (A -- 3Z = 6) nuclei between fluorine and magnesium ($^{33}$F, $^{36}$Ne, $^{39}$Na, and $^{42}$Mg) should be answered in the near future with the available increased intensities of the RIBF at RIKEN \cite{2007Yan01}. Beyond aluminum the dripline has most likely not been reached yet with the observation that $^{42}$Al is bound with respect to neutron emission \cite{2007Bau01}.

{\setlength{\LTleft}{0pt}
\setlength{\LTright}{0pt}


\setlength{\tabcolsep}{0.5\tabcolsep}

\renewcommand{\arraystretch}{1.0}
\renewcommand{\baselinestretch}{1.0}

\footnotesize 
\begin{longtable}{@{\extracolsep\fill}lllllll@{}}

\caption{Nuclides beyond the neutron dripline which have been demonstrated to be unbound or have not been reported yet.} \\
Z & Nuclide & Author & Year & Laboratory & Comments & Ref.\\
\hline\\
\endfirsthead\\
\caption[]{(continued)} \\
Z & Nuclide & Author & Year & Laboratory & Comments & Ref.\\
\hline\\
\endhead
4	 & 	$^{15}$Be	 & 	A. Spyrou	 & 	2011	 & 	MSU	 & 		 & 	\cite{2011Spy01}	  \\
	 & 	$^{16}$Be\footnote[1]{Discovered in 2012, see Section \ref{sec:2012}}	 & 	T. Baumann	 & 	2003	 & 	MSU	 & 		 & 	\cite{2003Bau01}	  \\
5	 & 	$^{20}$B	 & 	A. Ozawa	 & 	2003	 & 	RIKEN	 & 		 & 	\cite{2003Oza01}	  \\
	 & 	$^{21}$B	 & 	A. Ozawa	 & 	2003	 & 	RIKEN	 & 		 & 	\cite{2003Oza01}	  \\
6	 & 	$^{21}$C	 & 	M. Langevin	 & 	1985	 & 	GANIL	 & 		 & 	\cite{1985Lan01}	 	 \\
	 & 	$^{23}$C	 & 		 & 		 & 		 & 	not measured, but $^{21}$C unbound	 & 		 		 \\
7	 & 	$^{24}$N	 & 	H. Sakurai	 & 	1999	 & 	RIKEN	 & 		 & 	\cite{1999Sak01}		 \\
	 & 	$^{25}$N	 & 	H. Sakurai	 & 	1999	 & 	RIKEN	 & 		 & 	\cite{1999Sak01}	  \\
8	 & 	$^{26}$O\footnotemark[1]	 & 	D. Guillemaud-Mueller	 & 	1990	 & 	GANIL	 & 		 & 	\cite{1990Gui01}		 \\
	 & 	$^{27}$O	 & 	O. Tarasov	 & 	1997	 & 	GANIL	 & 		 & 	\cite{1997Tar01}		 \\
	 & 	$^{28}$O	 & 	O. Tarasov	 & 	1997	 & 	GANIL	 & 		 & 	\cite{1997Tar01}		 \\
9	 & 	$^{28}$F\footnotemark[1]	 & 	H. Sakurai	 & 	1999	 & 	RIKEN	 & 		 & 	\cite{1999Sak01}	 	 \\
	 & 	$^{30}$F	 & 	H. Sakurai	 & 	1999	 & 	RIKEN	 & 		 & 	\cite{1999Sak01}	 	 \\
	 & 	$^{32}$F	 & 		 & 		 & 		 & 	not measured, but $^{30}$F unbound	 & 		 		 \\
	 & 	$^{33}$F	 & 		 & 		 & 		 & 	potentially bound	 & 		 		 \\
10	 & 	$^{33}$Ne	 & 	M. Notani	 & 	2002	 & 	RIKEN	 & 		 & 	\cite{2002Not01}	 	 \\
	 & 	$^{35}$Ne	 & 		 & 		 & 		 & 	not measured, but $^{33}$Ne unbound	 & 		 		 \\
	 & 	$^{36}$Ne	 & 		 & 		 & 		 & 	potentially bound	 & 		 		 \\
11	 & 	$^{36}$Na	 & 	M. Notani	 & 	2002	 & 	RIKEN	 & 		 & 	\cite{2002Not01}	 	 \\
	 & 	$^{38}$Na	 & 		 & 		 & 		 & 	not measured, but $^{36}$Na unbound	 & 		 	 \\
	 & 	$^{39}$Na	 & 		 & 		 & 		 & 	potentially bound	 & 		 		 \\
12	 & 	$^{39}$Mg	 & 	M. Notani	 & 	2002	 & 	RIKEN	 & 		 & 	\cite{2002Not01}	 	 \\
	 & 	$^{41}$Mg	 & 		 & 		 & 		 & 	not measured, but $^{39}$Mg unbound	 & 		 		 \\
	 & 	$^{42}$Mg	 & 		 & 		 & 		 & 	potentially bound	 & 		  		 \\  \hline 	
\label{t:neutron-unbound}
\end{longtable}
}

\section{New Discoveries in 2012}
\label{sec:2012}

While in 2010 a record number of 110 new nuclei were reported \cite{2011Tho01}, only 7 additional new nuclei were discovered in 2011. The trend was again reversed in 2012 with the new identification of up to 67 nuclei. Kurcewicz et al. alone reported 59 new neutron-rich nuclei between neodymium and platinum \cite{2012Kur01}. These include \nuc{158}{Nd}, \nuc{178}{Tm}, and \nuc{181,812}{Yb} which had previously been reported only in conference proceedings (see Section \ref{subsec:proc}). Kurcewicz et al. reported the discovery of 60 new nuclides, however, \nuc{157}{Nd} was reported in a paper by Van Schelt et al. \cite{2012Van01} which had been submitted five months earlier. Van Schelt also measured  \nuc{155}{Pr} for the first time; both isotopes had previously been reported in a conference proceeding.
In addition, resonances in the light neutron-unbound nuclei  $^{16}$Be \cite{2012Spy01}, $^{26}$O \cite{2012Lun01} and $^{28}$F \cite{2012Chr01} were measured for the first time.

The remaining three nuclides, $^{95}$Cd, $^{97}$In, and $^{99}$Sn, bring up the discussion of what should be counted as a discovery. The particle identification plot in the recent publication by Hinke et al. exhibits clear evidence for the presence of $^{95}$Cd and $^{97}$In and a few events of $^{99}$Sn \cite{2012Hin01}. However, neither the text nor the figure caption mentions the discovery of these nuclides. In an earlier contribution to a conference proceeding Kr\"ucken reported the discovery of $^{95}$Cd and $^{97}$In, but not $^{99}$Sn, from the same experiment \cite{2008Kru01}.


In addition to these 66 nuclides another 6 new nuclides ($^{64}$Ti, $^{67}$V, $^{69}$Cr, $^{72}$Mn, $^{70}$Cr, and $^{75}$Fe) were reported  in a contribution to a conference proceeding \cite{2012Tar01}.



\section{Long Term Future}
\label{sec:future}

\begin{figure}
	\centering
	\includegraphics[scale=0.6]{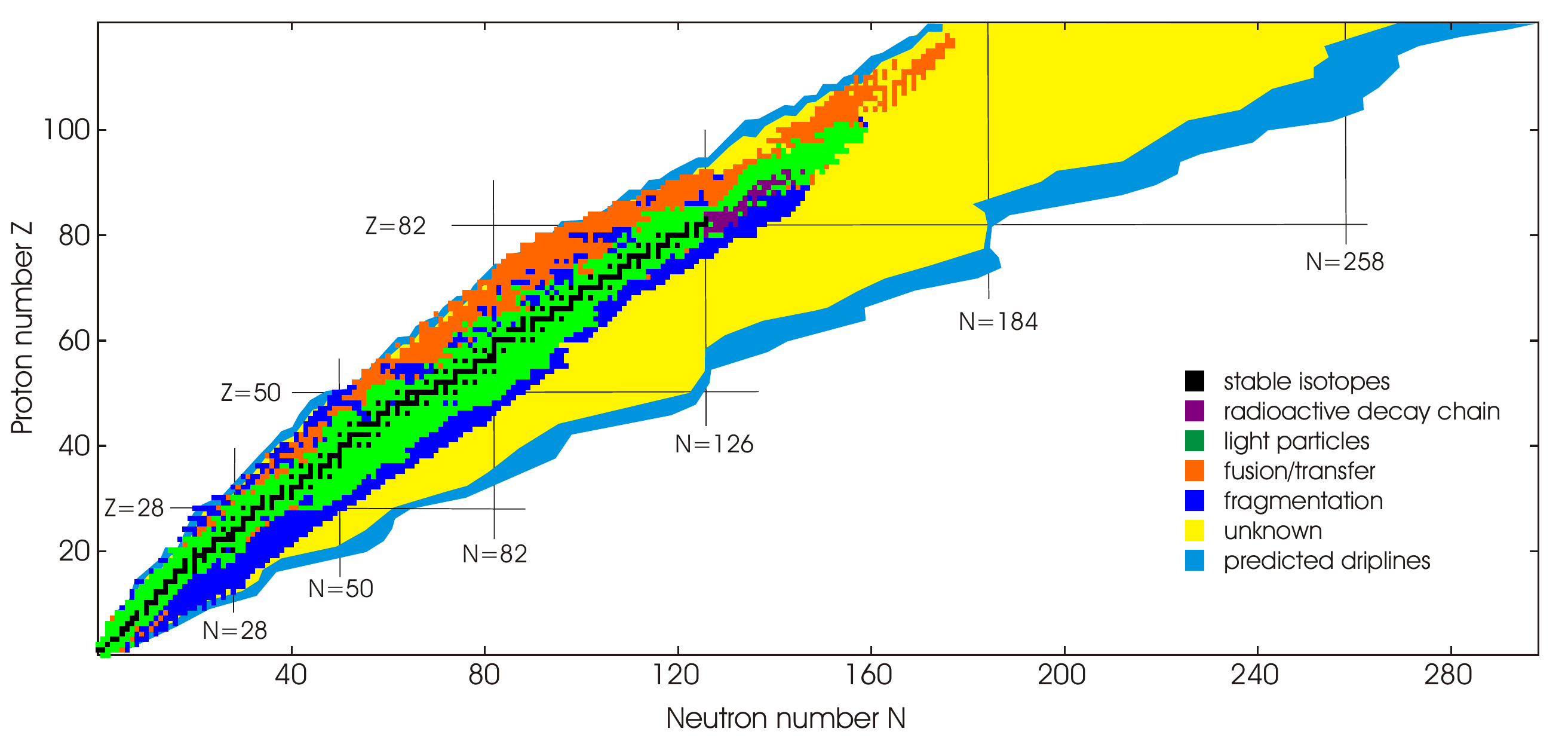}
	\caption{Chart of nuclides. Stable nuclides are shown in black. The other known nuclides are grouped according to the production mechanism of their discovery: radioactive decay chains (purple), light-particle induced reactions (green), fusion/transfer reactions (orange), and spallation or projectile fragmentation/fission (dark blue). Nuclides predicted to exist according to Ref. \cite{2012Erl01} are shown in yellow where the light-blue area shows the uncertainty of the driplines.}
\label{f:witek}
\end{figure}

Over 3000 different isotopes of 118 elements are presently known. In a recent article theoretical calculations revealed that about a total of 7000 bound nuclei should exist, thus more than double the nuclides presently known \cite{2012Erl01}. However, not all will ever been in reach as can be seen in Figure \ref{f:witek}. The figure shows the known nuclides first produced by light-particle reactions, fusion/evaporation reactions, and spallation/fragmentation which are shown in green, orange, and dark blue, respectively. Nuclides of the radioactive decay chains are shown in purple and stable nuclides in black. The yellow regions show unknown nuclides predicted by Ref. \cite{2012Erl01}. The light blue border corresponds to the uncertainty of the driplines in the calculations.

In the region of Z $>$ 82 and N $>$ 184 alone about 2000 nuclides will most probably never be created. If one conservatively adds another 500 along the neutron dripline in the region above Z $\sim$ 50 it can be estimated that another approximately 1500 nuclides (7000 predicted minus 3000 presently known minus 2500 out of reach) are still waiting to be discovered. In the 2004 review article on the limits of nuclear stability it was estimated that the Rare Isotope Accelerator (RIA) which had been proposed at the time would be able to produce about 100 new nuclides along the proton dripline below Z $\sim$ 82 \cite{2004Tho01}. Since then only about 20 of these nuclides have  been observed. Thus the next generation radioactive beam facilities (the Radioactive Ion-Beam Factory RIBF at RIKEN \cite{2010Sak01}, the Facility for Antiproton and Ion Research FAIR at GSI \cite{2009FAI01}, and the Facility for Rare Isotope Beams FRIB at MSU \cite{2010Bol01,2012Wei01}) should be able to produce approximately 80 new neutron-deficient nuclides. Equally critical for new discoveries at these facilities are the next generation fragment separators, BIG-RIPS \cite{2003Kub01,2008Ohn01}, the Super FRS \cite{2003Gei01}, and the FRIB fragment separator \cite{2011Ban01}, respectively.

Along the neutron dripline RIA was estimated to make another 400 nuclides below Z $\sim$ 50 \cite{2004Tho01} of which about 70 have been discovered in the meantime leaving about another 330 for the new facilities in the future.

The remaining unkown nuclides in the various regions of the nuclear chart have to be produced by different reaction mechanisms. Projectile fragmentation reactions will most likely be utilized to populate neutron-deficient nuclides below Z $\sim$ 50 and for nuclides above Z $\sim$ 82 fusion-evaporation reactions are the only possibility. The use of fusion-evaporation reactions with radioactive beams might be an alternative to reach nuclides which cannot be populated with stable target-beam combinations \cite{2004Tho01}. Neutron-deficient nuclides in the intermediate mass region (50 $<$ Z $<$ 82) have been produced so far by fusion-evaporation reactions, however, projectile fragmentation could be a viable alternative \cite{2000Sou01}.


New neutron-rich nuclides below Z $\sim$ 82 will most likely be only reachable by projectile fragmentation/fission reactions. The 2004 review predicted that the dripline would be reachable up to Z $\sim$ 30  \cite{2004Tho01}. If the dripline is as far away as estimated in the recent calculations \cite{2012Erl01} it could be that the dripline will not be reached beyond Z $\sim$ 16; at least not in the near future.

The search for new superheavy elements and therefore also new nuclei continues to rely on fusion-evaporation reactions \cite{2011Gat01,2012Mor01,2013Oga01}. However, recent calculations suggest that deep inelastic reactions or multi-nucleon transfer reactions on heavy radioactive targets (for example \nuc{248}{Cm}) might be a good choice to populate heavy neutron-rich nuclei \cite{2008Zag01,2011Zag01,2013Lov01}. The use of radioactive beams on radioactive targets could also be utilized for fusion-evaporation reactions in the future \cite{2013Lov01,2007Lov01}.

\section{Conclusion}
\label{sec:conclusion}

The quest for the discovery of nuclides that never have been made on Earth continues to be a strong motivation to advance nuclear science toward the understanding of nuclear forces and interactions. The discovery of a nuclide is the first necessary step to explore its properties. New discoveries have been closely linked to new technical developments of accelerators and detectors. In the future it will be critical to develop new techniques and methods in order to further expand the chart of nuclides .

The discovery potential is not yet limited by the number of undiscovered nuclides. About 1500 could still be created. This would correspond to about 90\% of all predicted nuclides below N $\sim$ 184 which should be sufficient to constrain theoretical models to reliably predict properties of all nuclides as well as the limit of existence.

\section*{Acknowledgements}

I would like to thank Ute Thoennessen for carefully proofreading the manuscript. This work was supported by the National Science Foundation under Grant No. PHY11-02511.

\bibliographystyle{iopart-num}
\bibliography{../isotope-discovery-references}

\end{document}